# Traceable precision pA direct current measurements with the ULCA

**Hansjörg Scherer[1], Gerd-Dietmar Willenberg[1], Dietmar Drung[2], Martin Götz[1], and Eckart Pesel[1]**

[1]Physikalisch-Technische Bundesanstalt, Bundesallee 100, 38116 Braunschweig, Germany

[2]Physikalisch-Technische Bundesanstalt, Abbestraße 2-12, 10587 Berlin-Charlottenburg, Germany

**Abstract**

A standard method for picoammeter calibrations is the capacitor charging technique, which allows generating traceable currents in the sub-nA range. However, its accuracy is limited by the ac-dc differences of the capacitances involved. The "Ultrastable Low-noise Current Amplifier" (ULCA) is a novel high-precision amperemeter for direct current measurements in the pA range, developed at PTB. Its amplifier stages, based on resistor networks and op-amps, can be calibrated traceably with a cryogenic current comparator (CCC) system. We compare the results from both independent calibration routes for two different ULCA prototypes. We find agreement between both methods at an uncertainty level below 10 µA/A, limited by the uncertainty of the currents generated with the capacitor charging method. The investigations confirm the superior performance of the new ULCA picoammeter.

# 1. Introduction

Small electrical currents are of increasing interest in the fields of fundamental and practical metrology. Recent advances in the field of single-electron transport devices offer ways for generating currents of the order of 100 pA with uncertainties of one part per million [1] or better [2] – [5]. This is expected to have fundamental impact on the future quantum-based realization of the SI unit ampere [6], [7]. Also for application purposes, e.g. for applications in dosimetry and semiconductor industry, there is increasing need for sub-nA current measurements, requiring the calibration of picoamperemeters (picoammeters). In order to underpin their calibration and measurement capabilities in this field, lately thirteen National Measurement Institutes (NMIs) had evaluated their corresponding calibration methods by an international comparison for the first time [8].

In this paper we present the results of a comparison of complementary state-of-the-art techniques for the traceable generation and measurement of direct currents in the pA range at highest accuracy. The capacitor charging method was used for generating sub-nA currents, serving as calibration currents applied to two different prototypes of a novel picoammeter instrument presently under development at PTB. A new cryogenic current comparator (CCC) setup at PTB was used for independent calibrations of the picoammeter prototypes. The comparison of the results from both calibration routes gives insight into the limitations of the capacitor charging method for current generation, and confirms the superior performance of the new picoammeter instrument.

# 2. Background and methods

The capacitor charging method [9] is a well-established primary standard technique for the generation of small direct currents, commonly used by NMIs for calibrations in the sub-nA current regime [8]. With this method, a current is generated by charging or discharging a high-precision, usually gas-filled, capacitor of capacitance $C$ by applying a voltage ramp, i.e. a voltage $V$ linearly increasing or decreasing with slope $dV/dt$. The resulting current $I = C \cdot dV/dt$ is, thus, traced back to the SI units volt, second and farad (see Fig. 1 for the schematics of the setup).

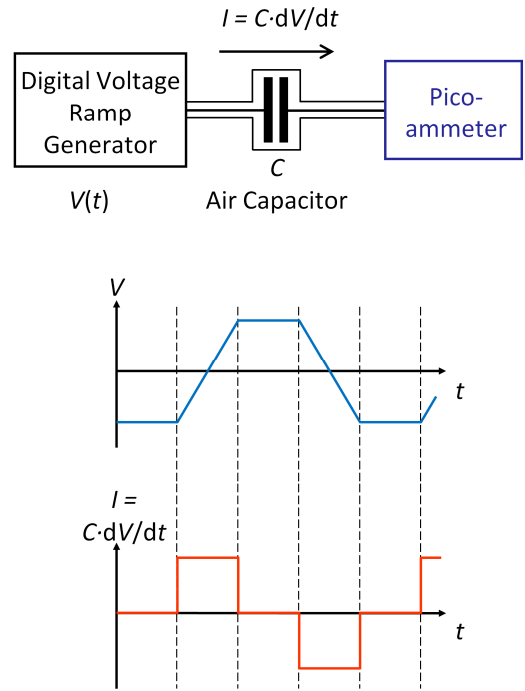

**Figure 1:** Schematics of sub-nA current generation using the capacitor charging method [9]. A voltage ramp is used to periodically charge and discharge a precision capacitor. The output current from the capacitor terminal, traceable to volt, second and farad, is used to calibrate a picoammeter.

The frequency dependence of $C$ was found crucial for the accuracy of the method [10]. At best, 100 pA currents can be generated with relative uncertainties of about 10 µA/A [8], [11].

The precision of sub-nA current measurements with conventional commercial picoammeters, even following calibration, is nowadays limited to about 10 µA/A due to amplifier gain drifts. This is insufficient for future fundamental and practical metrological applications. Up to now, only one rather sophisticated setup was reported that achieved 1 µA/A accuracy for a measurement current in the range of 100 pA [1]. Lower uncertainties may be achieved using more complex measurement systems based on CCCs with high winding ratios [12], [13]. However, the SQUID null detector typically involved in these setups may cause possible nonlinear effects in the regime of very low input current (flux) levels. Therefore, recently doubts have arisen concerning systematic imponderabilities in pA current measurements performed with CCC systems [15], [16].

To improve this situation, PTB pursues the development of a new picoammeter: the "Ultrastable Low-noise Current Amplifier" (ULCA) is a fully non-cryogenic single-box instrument based on specially designed op-amps and resistor networks [7], [14] − [16]. The amplifier's transfer coefficient, its transresistance $A_{TR} = U_{out} / I_{in}$, is extremely stable versus temperature fluctuations and input current amplitude [15], [16]. It can be calibrated traceable to the quantum Hall resistance (QHR) by using a CCC at current levels in the nA range. For the experiments reported in this paper we used two ULCA prototypes from different stages of development and performance levels.

## 3. Experimental details

### 3.1. ULCA prototypes and CCC-based calibration

The single-box ULCA instruments are based on a double-stage amplifier concept: the first stage, realized by specially designed resistor networks, amplifies the input current with a gain factor $G_I = 1000$ (nominally); the second stage realizes a current-to-voltage conversion via a resistor ($R_{I-V}$). The total transfer coefficient of the ULCA, the effective transresistance, is thus given by $A_{TR} = G_I \cdot R_{I-V}$, and it can be calibrated traceably to the quantum Hall resistance (QHR) with a CCC at current levels in the nA range. More technical details about the ULCA concept and performance characteristics are given in [15] and [16].

The first ULCA prototype (ULCA-0) had an effective nominal transresistance $A_{TR-0}$ of 50 MΩ (i.e. $R_{I-V}$ = 50 kΩ). The temperature coefficient for the channel used was +1.8 μΩ/Ω per Kelvin, and the effective input current noise level was about 4 fA/√Hz, constant ("white") in the frequency range between 0.01 Hz and 0.3 Hz [7]. The second instrument ULCA-1 with $R_{I-V}$ = 1 MΩ was a significantly improved prototype from the latest development stage [14] − [16] (photo see Fig. 2). Its much higher effective transresistance $A_{TR-1}$ = 1 GΩ had a very low temperature coefficient of only -0.22 μΩ/Ω per Kelvin. The effective input current noise level was only 2.4 fA/√Hz, with a 1/$f$ corner below 1 mHz.

During all measurements, the temperature inside the ULCA instruments was monitored by using internal temperature sensors. Due to the special low power consuming design of both ULCA prototypes, the difference between room temperature and their effective internal temperature typically was less than 0.3 K.

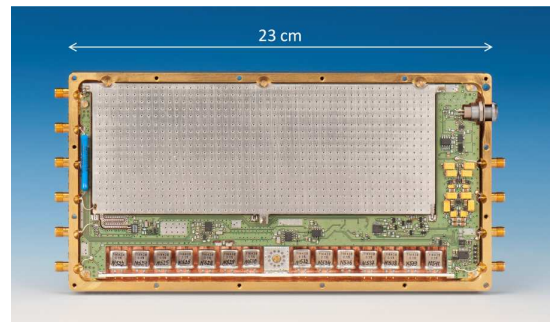

**Figure 2:** Photo of the ULCA-1 picoammeter prototype with a total effective transresistance of nominally 1 GΩ [15], [16]. The electronic circuit is situated in a solid copper box for thermal stabilization (upper enclosure opened here). The 3 GΩ / 3 MΩ resistor network for the 1:1000 current amplification stage, made from thin-film chip resistors, is located under the silvery guarding board. The resistor array visible at the bottom corner of the box is the current-to-voltage converter stage, made from metal-foil precision resistors with a total resistance $R_{I-V}$ = 1 MΩ. All connections to inputs and outputs of the instrument are made with SMA connectors.

For the CCC-based calibrations of $A_{TR}$ different procedures were applied for both prototypes: the calibration of ULCA-0 was performed with 12-bit CCC (PTB's standard CCC system for resistance calibrations) at an input current of about 3 nA. For this calibration, the transresistance $A_{TR-0}$ of 50 MΩ was directly compared to a precision standard resistor of about 12.9 kΩ resistance. This calibration was traceable to the QHR with a total relative uncertainty of 1 µΩ/Ω. An initial CCC-based calibration performed one year before gave a calibration result for $A_{TR\text{-}0}$ differing only by about 6 µΩ/Ω, which confirms the excellent stability of the ULCA.

Because of its considerably larger nominal transimpedance $A_{TR\text{-}1}$ of 1 GΩ, the improved ULCA-1 prototype was calibrated in two steps using the new binary "14-bit" CCC developed at PTB [17], [18]: in the first step, the current amplification factor $G_I$ from the first amplifier stage (a resistor network made from 3 GΩ and 3 MΩ resistors for nominally 1000-fold current amplification [14] − [16]) was calibrated with the CCC alone, i.e. without involving an external standard resistor. This calibration was done using CCC windings with a turn ratio of about 16 000:16 and currents of 13 nA and 13 µA, respectively. In the next step, the resistor of the current-to-voltage converter stage ($R_{I\text{-}V}$ = 1 MΩ) was calibrated in comparison to a precision standard resistor of about 12.9 kΩ at a current of 500 nA. This procedure allowed the calibration of the transresistance $A_{TR\text{-}1}$ with a total relative uncertainty of 0.06 µΩ/Ω traceable to the QHR. The drift of $A_{TR\text{-}1}$ observed over a period of six days was less than 0.7 µΩ/Ω.

*3.2. Calibration current generation with the capacitor charging method*

The measurement setup was situated in a temperature-controlled electrically shielded room at about $T = 23$ °C. Humidity and air pressure were not stabilized.

For the capacitor charging we used a digital voltage ramp generator, developed for picoammeter calibrations with the capacitor charging method at PTB [19]. The computer-controlled instrument is capable of generating highly linear voltage ramps running between -10 V and +10 V with highly stable and linear slope, adjustable between 1 mV/s and 1 V/s. The ramp generator output was monitored using a calibrated 8 ½-digit multimeter (DMM), triggered by a stabilized precision time base.

The ramp generator output was connected to commercial capacitors of type GR1404 (hermetically sealed in dry nitrogen dielectric) with nominal capacitances $C = 1$ nF. The capacitors were calibrated by using a commercial precision capacitance ac bridge at the operation frequency of 1 kHz. For some measurements two nominally identical capacitor specimens were used in order to test the influence their possibly unequal ac-dc differences.

The calibration currents ranging from 50 pA to 500 pA (corresponding to the 1 nF capacitor and voltage slopes between 50 mV/s and 500 mV/s) were fed to the ULCA input via a short piece of low noise coaxial cable. Current generation was performed in subsequent cycles, each of them consisting of the four phases i) zero current, ii) positive current, iii) zero current, and iv) negative current (cf. Fig. 1). The duration of each phase ranged between 40 s and 200 s. The ULCA output voltage $U_{out}$ was sampled by an 8 ½-digit DMM, calibrated with a Josephson voltage standard system. Data acquisition was performed over periods between eight and 66 hours of duration, depending on the current value. After this acquisition time, statistical uncertainty contributions typically were at a level of two parts per million or below. The accuracy of this calibration method corresponded to total relative uncertainties of about 11 µA/A, dominated by the systematic uncertainty contribution of 10 µA/A assigned to the effect of the capacitor ac-dc difference [10].

## 4. Results and discussion

The experimental results are summarized in Fig. 3. The two graphs show the transresistance figures $A_{TR}$ for both ULCA prototypes determined by the two independent ULCA calibration methods: first, the capacitor charging method (red data points) and, second, the CCC calibrations (blue data points). All error bars correspond to standard uncertainties (coverage factor $k = 1$). Uncertainty bars for the capacitor charging method are shown without (small red bars) and including (larger red bars) the 10 µΩ/Ω contribution from the frequency dependency of the capacitors [10].

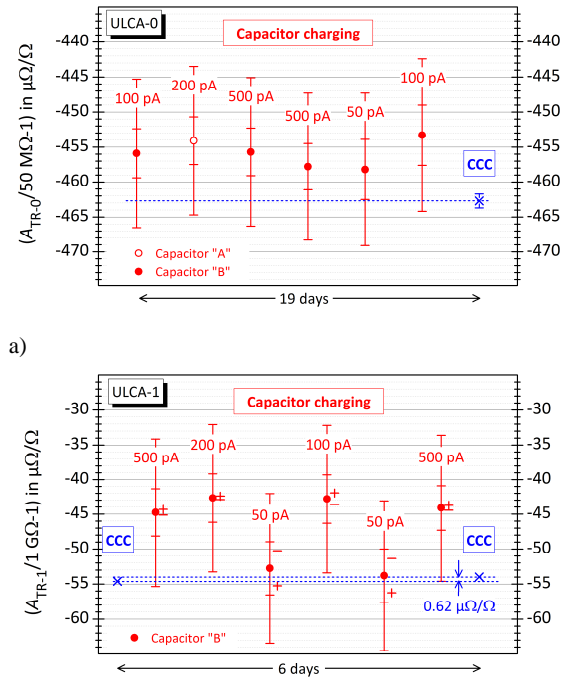

a)

b)

**Figure 3:** Comparison of the two ULCA calibration methods via capacitor charging (red dots) and CCC (blue crosses) for two ULCA prototypes. The plots show the relative differences of the ULCA transresistance values $A_{TR}$ to their nominal values. In each graph the data are plotted in the chronological order of the measurements. All error bars correspond to standard uncertainties ($k = 1$). Uncertainty bars for the capacitor charging method are shown without (small red bars) and including (larger red bars) the contribution of ten parts per million due to frequency dependency of the capacitors. Panel a) shows the results for ULCA-0 (all data are related to a temperature of 24.7 °C). The CCC calibration of $A_{TR-0}$ was performed with a total relative uncertainty of one part per million. Panel b) shows the results for ULCA-1 (all data in are related to 23.0°C). The total relative uncertainty from the CCC calibration of $A_{TR-1}$ was 0.06 parts per million (not visible in the plot). The drift of $A_{TR-1}$ observed over the measurement period of six days was 0.62 µΩ/Ω. +/- symbols in panel b) show $A_{TR-1}$ values measured for positive and negative current directions for each calibration current value. See text for more details and discussion.

In each graph the data are plotted in the chronological order of the measurements.

Small corrections considering the temperature coefficients for $A_{TR-0}$ and $A_{TR-1}$ were applied to the data.

Panel a) shows the results from the measurements with the ULCA-0 ($A_{TR-0} = 50$ MΩ nominally), taken over a period of 19 days. The results from the CCC calibration agree to the currents generated with the capacitor charging method within the total standard uncertainty of 11 µA/A attributed to the values from the latter method. However, the values determined with the capacitor charging method appear slightly (about 5 µΩ/Ω to 10 µΩ/Ω), but systematically higher than the value determined from the CCC calibration. This is attributed to the frequency dependence (ac-dc difference) of the capacitors used for the current generation method via capacitor charging. To shed more light on this effect, for some measurements two different, but nominally identical capacitor specimens were used (GR1404 "A" and "B", both with $C = 1$ nF). An example for a measurement at $I = 200$ pA using GR1404 "A" is shown in Fig. 3 a. These investigations, however, showed no significant difference between the individual capacitor specimens.

The effect of the frequency dependence of different capacitor specimens used for current generation with the capacitor charging method was investigated in [10], in particular also regarding the PTB capacitor specimens GR1404 "A" and "B". It was found that their capacitance values increased when the measuring frequency was lowered from the 1 kHz ("ac" operation frequency of the capacitance bridge used for the capacitor calibration) down to the range of mHz frequencies (i.e. the effectively relevant voltage ramp frequencies for the capacitor charging method at "dc"). Although GR1404 "A" and "B" had shown the smallest frequency dependencies among all capacitors investigated in [10], their relative ac-dc differences had been found to be five to ten µF/F; however, the uncertainties attributed to these figures had been of the order of 10 µF/F. Therefore, the ac-dc difference effect of the GR1404 "A" and "B" capacitors used for the charging method in the PTB setup was considered conservatively by adding an uncertainty contribution of 10 µF/F to the uncertainty budget [8], [11]. The results shown in Fig. 3 a in this paper confirm the results on GR1404 "A" and "B" from [10]. The finding that the $A_{TR-0}$ values determined with the capacitor charging method all are few µΩ/Ω larger than the CCC calibration values is consistent with a negative frequency coefficient also found in [10]: an effectively larger capacitance at "dc" results in current generator output currents that are larger than the nominal value assumed. Such currents fed to the ULCA cause accordingly larger output voltages and, thus, larger calibration values for $A_{TR-0}$. Also, the magnitude of about 5 µΩ/Ω to 10 µΩ/Ω of the ac-dc capacitance difference effects is consistent with [10].

The findings made in the measurements using the prototype ULCA-0 were qualitatively confirmed by the measurements with the improved prototype ULCA-1. The results from this six-day measurement run, performed about five months after the measurements with ULCA-0, are shown in Fig. 3 b. Here, the $A_{TR-1}$ values from the capacitor charging technique (using capacitor specimen "B") also agree with the values determined from the CCC calibrations within their standard uncertainty of about 11 µΩ/Ω. Also, as before, for $I = 100$ pA to 500 pA the capacitor charging method resulted in systematically higher calibration values (9 µΩ/Ω to 11 µΩ/Ω) compared with the CCC calibrations. The calibration values for $I = 50$ pA, however, were found about 8 µΩ/Ω to 10 µΩ/Ω lower than the values measured for $I = 100$ pA to 500 pA. This behaviour was reproduced by repeating the 50 pA measurement (Fig. 3 b). In both measurements with 50 pA, the $A_{TR-1}$ values exhibited significant and reproducible differences for positive and negative current directions (cf. the plus and minus symbols for the $A_{TR-1}$ values obtained for each current direction and each calibration current in panel b) of Fig. 3). Experimental results from independent high-precision characterization measurements and the design principle of the ULCA, including op-amps with open loop gains of $10^9$ and more indicate that such asymmetry errors on this magnitude scale are not expected to be caused by the ULCA instruments [16]. Although the ominous effects found in this measurement run at 50 pA are not completely understood yet and subject of ongoing instigations, we attribute them to the current generation method.

## 5. Conclusion and outlook

Two independent and traceable methods were applied for the calibration of prototypes of the new ULCA picoammeter. The calibration results from both methods are in agreement. Since the capacitor charging method is a primary standard technique for the generation of sub-nA currents, the ULCA including the CCC-based calibration routine is considered validated successfully. In particular, the accuracy and the linearity of the ULCA prototypes were confirmed over a large current range from 50 pA up to a few nA.

The limitations of the calibration comparison presented in this paper were given by the comparably large uncertainty of the capacitor charging method. The excellent accuracy and stability of the ULCA picoammeters, following CCC-based calibration, allowed investigating systematic effects and uncertainties of the standard sub-nA current generation technique. The uncertainty contribution arising from the capacitors' frequency dependence (ac-dc difference) involved in our current generation setup was found to be about five to ten parts per million. This quantitatively confirmed earlier investigations presented in [10], and in particular also the total uncertainty of about ten parts per million attributed to the capacitor charging method at PTB up to present [11].

In conclusion, the advantages of the ULCA concept are given by the excellent stability of its amplification coefficient $A_{TR}$ together with the fact that it can be calibrated traceable to electrical quantum standards with a suitable CCC. The precision of the ULCA is excelling the hitherto state-of-the-art sub-nA current generation method by more than one order of magnitude, and it is about two orders of magnitude better than the precision of the best commercially available instruments. This now allows measuring electrical currents up to about 10 nA with unparalleled accuracy, and we expect the ULCA to set new standards for small-current calibration applications.

The latest ULCA prototype version has a very low effective input current noise of about 2.4 fA/√Hz, and its amplification coefficient can be calibrated using our 14-bit CCC with a relative uncertainty well below one part in $10^7$. With these features, the ULCA is capable of measuring a 100 pA current with an uncertainty of one part in $10^7$ in a measurement time of about 10 hours (with periodic current reversals to eliminate flicker noise effects, and considering 20% data rejection for settling after the reversals). This superior performance not only offers benefits for calibration applications, but also has implications for fundamental metrology research, for instance in the field of single-electron pumps. Recent developments suggest the use of advanced combinations of "high current" (i.e. generating 100 pA or more) single-electron pumps in combination with single-electron detectors for a future quantum-based realisation of the ampere via exploiting the basic definition of electric current $I = e \cdot f$. Among the most promising candidates for suitable single-electron pumps are devices based on dynamic quantum dots fabricated from semiconductor systems [1] − [5], [7], [20] − [23]. The theoretical understanding of the relevant transport mechanisms in these devices is rather developed (for details see [23] and references therein). However, the identification of the underlying dominant dynamical effects in different transport regimes requires the precise analysis of curvature shapes of their current-voltage characteristics in the pumping regime. The validation and optimization of these pump devices also requires absolute measurements of their quantized current plateaux levels at about 100 pA. Both tasks need absolute current measurements of currents in the 100 pA range, accurate on a precision level of one part per million or better. The handy ULCA offers a practical way to perform such measurements in the future, and is therefore expected to be a relevant cornerstone also in the field of single-electron research and metrology.

## Acknowledgements

We thank Franz Josef Ahlers for support with the measurement software, Christian Krause, Michael Piepenhagen and Monique Klemm for assembling electronics boards, and Ralf Behr and Susanne Gruber for the voltmeter calibrations.

This work was done within Joint Research Project "Qu-Ampere" (SIB07) supported by the European Metrology Research Programme (EMRP). The EMRP is jointly funded by the EMRP participating countries within EURAMET and the European Union.